\UseRawInputEncoding
\documentclass[aip,amsmath,amssymb,reprint,floatfix]{revtex4-1}
\usepackage{graphicx}
\graphicspath{ {./imgs/} }
\usepackage{soul}
\usepackage{xcolor}
\usepackage{afterpage}

\sethlcolor{white}

\makeatletter
\DeclareRobustCommand\citenum
   {\begingroup
     \NAT@swatrue\let\NAT@ctype\z@\NAT@parfalse\let\textsuperscript\relax
     \NAT@citexnum[][]}
\makeatother

\begin{document}


\title{Efficient Grand Canonical Global Optimization with On-the-fly-trained Machine-learning Interatomic Potentials} 



\author{Jon Eunan Quinlivan Dominguez}
\affiliation{Departament de Ci\`encia de Materials i Qu\'{\i}mica F\'{\i}sica and Institut de Qu\'{\i}mica Te\`{o}rica i Computacional (IQTCUB), Universitat de Barcelona, c/ Mart\'{\i} i Franqu\`es 1, 08028 Barcelona, Spain}

\author{Mads-Peter Verner Christiansen}
\affiliation{Center for Interstellar Catalysis, Department of Physics and Astronomy,
Aarhus University, DK-8000 Aarhus C, Denmark}

\author{Konstantin M. Neyman}
\affiliation{Departament de Ci\`encia de Materials i Qu\'{\i}mica F\'{\i}sica and Institut de Qu\'{\i}mica Te\`{o}rica i Computacional (IQTCUB), Universitat de Barcelona, c/ Mart\'{\i} i Franqu\`es 1, 08028 Barcelona, Spain}
\affiliation{ICREA (Instituci\'{o} Catalana de Recerca i Estudis Avan\c{c}ats), Pg. Llu\'{\i}s Companys 23, 08010 Barcelona, Spain}

\author{Bj$\o$rk Hammer}
\affiliation{Center for Interstellar Catalysis, Department of Physics and Astronomy,
Aarhus University, DK-8000 Aarhus C, Denmark}

\author{Albert Bruix}
\email{abruix@ub.edu}
\affiliation{Departament de Ci\`encia de Materials i Qu\'{\i}mica F\'{\i}sica and Institut de Qu\'{\i}mica Te\`{o}rica i Computacional (IQTCUB), Universitat de Barcelona, c/ Mart\'{\i} i Franqu\`es 1, 08028 Barcelona, Spain}



\date{\today}

\begin{abstract}
The characterization of nanostructured materials under reactive environments is challenging due to the complexity of the structural motifs involved and their chemical transformations. Global optimization approaches allow predicting stable structures for targeted materials but addressing the configurational and compositional search spaces is both computationally demanding and inefficient, especially when first principles calculations are required. In this work, we implement and evaluate a computationally efficient grand canonical global optimization algorithm able to identify stable structures and chemical states of targeted systems under given reaction conditions (e.g. reactant pressure and temperature). The algorithm leverages an on-the-fly trained machine-learning interatomic potential based on sparse Gaussian Process Regression and the smooth overlap of atomic positions descriptor to reduce the number of first principles energy evaluations carried out during global optimization searches. The \textit{ab initio} thermodynamics framework is incorporated to approximate the Gibbs energy of evaluated candidates, performing environment-aware optimizations over multiple stoichiometries. We demonstrate the computational performance of this approach and its ability to reproduce some literature examples.
\end{abstract}

\pacs{}

\maketitle 

\section{Introduction}
The characterization of nanostructured materials is challenging due to their significant structural complexity. Furthermore, their size, structure, and chemical state not only depend on preparation conditions, but are also often affected by the interaction with reactive environments. This is particularly relevant for catalytic materials, for which \textit{in situ} or \textit{operando} spectroscopic techniques\cite{catal-Newton2008,catal-Kalz2016,catal-Hansen2002} have revealed a highly dynamic response upon exposure to reaction conditions. The understanding of structure-activity relationships in catalysis research is therefore precluded by these transformations, which are hard to characterize at the atomic level due to the loss in resolution of spectroscopic techniques at elevated temperature and reactant pressure. Similarly, the computational characterization of the properties of catalytic materials based on first principles approaches is reliable only if the structural models used are representative of the geometries and chemical states present at the targeted reaction conditions. We note that by chemical state of the system we refer to the number of atoms of each element it contains, i.e., its stoichiometry or chemical formula. 

Building such representative computational structural models typically involves searching for stable structures (or the most stable one, i.e. the global minimum) for a given set of relevant stoichiometries. However, this is challenging due to the large number of atoms (and species) involved and to the resulting high dimensionality of the composition and configurational spaces \cite{catal-Grajciar2018, catal-Lykhach2015}. To overcome the limitations of heuristic solutions to this problem, several automated \textit{Global Optimization} methods\cite{catal-Johnston2002} such as Monte Carlo sampling and Evolutionary algorithms\cite{catal-Wang2019,ga-Vilhelmsen2012,ga-Vilhelmsen2014,gaml-Vilhelmsen2012,catal-Liu2018,catal-Zhai2017,catal-Zandkarimi2019,goga-Reichenbach2019} have been developed in the last few decades. However, a thorough search for the Global Minimum for a single stoichiometry typically requires hundreds or thousands of energy evaluations. In order to consider various chemical states, this search has to be extended to multiple stoichiometries, increasing the number of required evaluations with each additional adsorbate or element considered\cite{pt3ox-QuinlivanDomnguez2022}.

To reduce the number of required first principles energy calculations to globally optimize a structure, the use of machine learning interatomic potentials (MLIPs) such as supervised regression models has been incorporated into global optimization algorithms in various novel approaches\cite{nn-Behler2007,nn-Lubbers2018,nn-Schtt2017,nn-Schtt2018}. MLIPs are typically trained on large databases of atomic structures, and their calculated energies can approach the accuracy of the level of theory used for the creation of the database\cite{mlatomic-Xie2023,mlmace-Kovcs2023}. This allows one to partially or completely replace expensive \textit{ab initio} energy evaluations in global optimization algorithms, allowing a more thorough or computationally efficient exploration of the potential energy surface (PES)\cite{gofee-Bisbo2020,gofee-Bisbo2022,mlip-Deringer2018,mlip-Behler2015,mlip-Jinnouchi2019,mlip-Li2015,mlip-Lin2020,mlip-Loeffler2020,mlip-Schmitz2020,mlip-Timmermann2021,mlip-Xu2021,agox-Christiansen2022,agoxlgpr-Rnne2022}. However, if no existing databases or reliable pre-trained models are available, the computational cost associated with the construction of databases for model training can trump the savings achieved by using the trained potential in a given application. 

Many MLIPs in fact require a large number of training examples to reach the desired accuracy, particularly those based on neural-network architectures \cite{nn-Behler2007,nn-Lubbers2018,nn-Schtt2017,nn-Schtt2018}. In turn, models based on Gaussian Process Regression (GPR)\cite{gap-Bartk2010,gap-Bartk2017,gap-Chmiela2017,gap-Deringer2018,gap-Klawohn2023} are typically more data-efficient and can reach chemical accuracy with a relatively small number of training examples. This, alongside the simpler fitting procedure compared to other MLIPs, makes GPRs particularly suited for global optimization algorithms based on first principles calculations assisted by an on-the-fly trained MLIP. Seminal examples of such novel approaches are the GOFEE algorithm \cite{gofee-Bisbo2020,gofee-Bisbo2022} and its subsequent improvements in the AGOX platform \cite{agox-Christiansen2022}, where the targeted level of theory is used to perform single-point calculations of carefully selected candidates, and a GPR MLIP is actively trained and used to carry out local relaxations and provide uncertainty measures that guide the exploration of the optimization algorithm. Remarkably, these approaches require orders of magnitude fewer first principles energy evaluations than those not exploiting MLIPs to find the global minimum of targeted systems. 

Despite the high computational efficiency of these machine-learning assisted global optimization approaches in finding the global minimum, most implementations are designed to optimize the structure for a system with a fixed number of atoms, i.e., a single stoichiometry at a time. Thus, in order to assess the relative stability of structures with different chemical states, one needs to find the global minimum for each considered stoichiometry and compare their Gibbs energy of formation in a posterior  \textit{ab initio} thermodynamics\cite{thermo-Reuter2003,thermo-Reuter2005,thermo-Reuter2016} analysis. This approach has been successfully applied in numerous studies to predict stable structures and oxidation states of supported metal and oxide particles at catalytically relevant conditions\cite{goga-Reichenbach2019,pt3ox-QuinlivanDomnguez2022}, but it suffers from some drawbacks. Namely, the number of individual stoichiometries that need to be considered scales linearly with the system size and exponentially with the number of constituent elements (or adsorbates), the abundance of which is optimized during a run. For example, optimizing the structure and chemical state of a Pt$_6$O$_x$ cluster involved identifying the global minimum for each stoichiometry with $x$ from 0 to 12 (13 stoichiometries) \cite{pt3ox-QuinlivanDomnguez2022}. Considering the possibility of adsorbing up to one CO molecule on each Pt atom leading to Pt$_6$O$_x$(CO)$_y$ systems would involve finding the global minima for 78 stoichiometries. Not only is this scaling critical to computational efficiency and feasibility, but many of the evaluated stoichiometries in any such search are very unstable and therefore irrelevant.

Grand canonical global optimization approaches can address these drawbacks by evaluating the Gibbs energy of formation at the conditions of interest for each candidate directly during search execution \cite{gcgo-BonillaPetriciolet2011,gcgo-Rangaiah2001,gcgo-Teh2002,gcgoml-Zhang2025,gcgo-bs1,gcgo-bs2,gcgo-bs3}. Both configurational and compositional spaces can be simultaneously explored to identify the most stable structure and chemical state. Such algorithms have been implemented and applied in previous works, but exploration of the chemical space increases the already high computational demands of global optimization searches in such non ML-assisted approaches \cite{gcgo-Calvo2016,gcgo-Rangaiah2001,gcgo-Calvo2016,gcgo-BonillaPetriciolet2010EvaluationOS}.

In this work, we leverage the capacity of MLIPs to accelerate global optimization algorithms to develop and implement a machine-learning assisted grand canonical evolutionary algorithm. This algorithm is implemented as a set of customizable components in the Atomistic Global Optimization X (AGOX) python library\cite{agox-Christiansen2022}. The algorithm relies on the smooth overlap of atomic positions (SOAP)\cite{locald-Bartk2013} description of the atomic environments and a local GPR model as implemented in the AGOX framework\cite{agoxlgpr-Rnne2022} to train, on-the-fly, a size-extensive MLIP suitable for multiple stoichiometries. As in previous single stoichiometry (or canonical) evolutionary algorithms assisted by machine-learning, the on-the-fly trained MLIP is used to carry out structural relaxations during the global optimization search, whereas first principles calculations are used only for single-point evaluations of candidates selected based on stability and uncertainty criteria. 

The performance of the algorithm is illustrated in structure and chemical state searches of various example systems. Specifically, we demonstrate the capacity to optimize free-standing Ir$_3$O$_x$ clusters, Pt$_3$O$_x$ clusters supported on the CeO$_2$(111) surface, and surface oxide reconstructions on the Pd(100) surface.

\section{Methods and Computational Details}
\subsection{Implementation of the grand canonical global optimization algorithm}

\begin{figure*}[ht]
    \centering
    \includegraphics[width=0.9\textwidth]{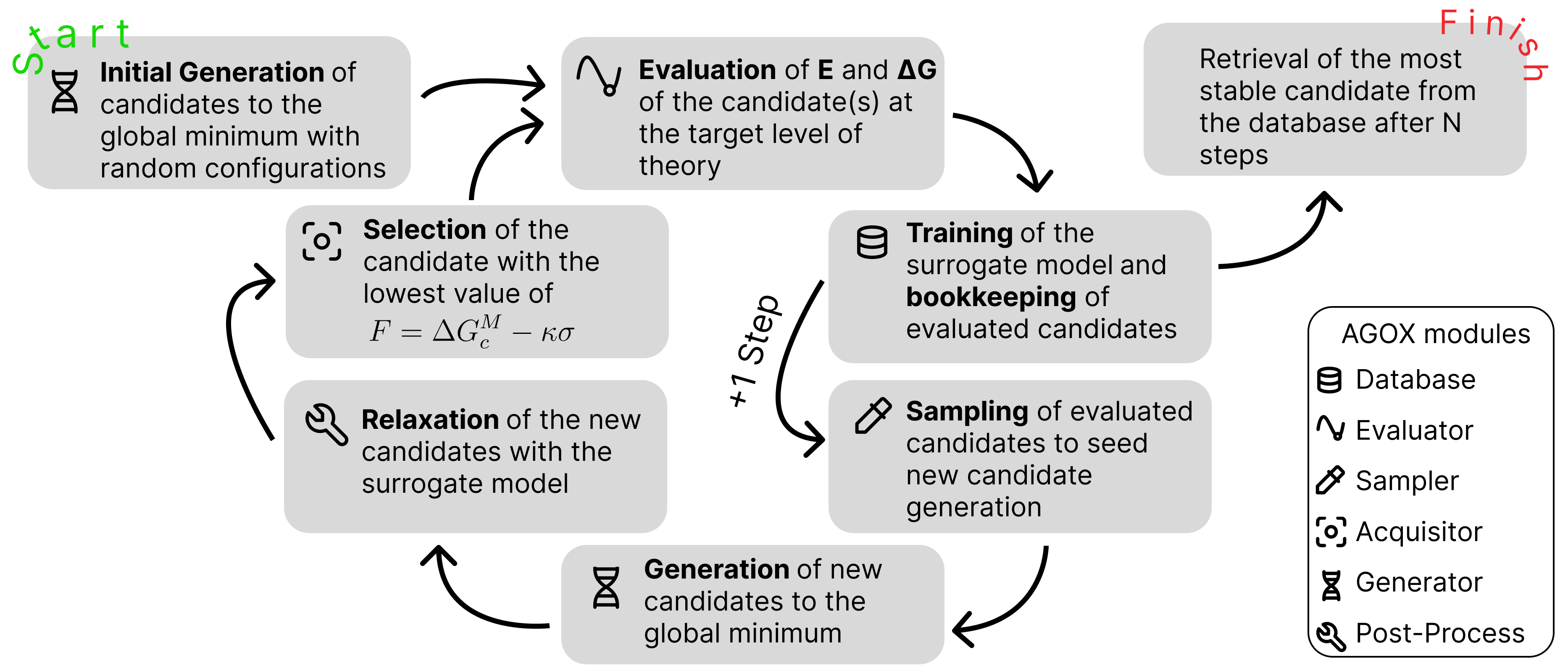}
    \caption{Execution diagram of the grand canonical global optimization algorithm as implemented with the AGOX modules.The diagram shows an iteration of the algorithm consisting of 1) the initial generation of a set of randomly generated candidates with the Generator module; 2) evaluation of the energies of structures of this set at the target level of theory with the Evaluator module; 3) training of the surrogate energy model and book-keeping within de Database module; 4) selection by the Sampler module of $N$ candidates from the database based on their fitness; 5) generation of new candidates with the Generator module taking selected candidates as seed; 6) relaxation of the new candidates with the surrogate energy model within the Post-process module; 7) selection by the Acquisitor module of the fittest new candidate according to the surrogate energy model; 8) evaluation of the selected candidate at the target level of theory; back to step number 3). }
    \label{gcgo-diagram}
\end{figure*}

The implemented grand canonical global optimization (GCGO) algorithm aims to find the most stable structure and chemical state of a target system exposed to given reaction conditions.  This is achieved by combining advances in molecular structural optimization algorithms and the \textit{ab initio} thermodynamics framework \cite{thermo-Reuter2003,thermo-Reuter2005,thermo-Reuter2016}. Specifically, the GCGO algorithm performs the search following the same strategy as the GOFEE method \cite{gofee-Bisbo2020,gofee-Bisbo2022}, but adapted to consider multiple stoichiometries whose Gibbs energy is evaluated at the targeted conditions. Both the GCGO and single-stoichiometry algorithms share the same execution diagram shown in Fig. \mbox{\ref{gcgo-diagram}}, which proceeds as follows:
\begin{enumerate}
    \item Initial generation of a set of randomly generated candidates with the \textbf{Generator} module
    \item Evaluation of the energies of structures of this set at the target level of theory with the \textbf{Evaluator} module
    \item Training of the surrogate energy model and book-keeping within de \textbf{Database} module
    \item Selection by the \textbf{Sampler} module of $N$ candidates from the database based on their fitness
    \item Generation of new candidates with the \textbf{Generator} module taking selected candidates as seed
    \item Relaxation of the new candidates with the surrogate energy model within the Post-process module
    \item Selection by the \textbf{Acquisitor} module of the fittest new candidate according to the surrogate energy model
    \item Evaluation of the selected candidate at the target level of theory with the \textbf{Evaluator} module
    \item End iteration and back to step number 3
\end{enumerate}

The two algorithms thus begin by producing a user-selected number of candidates with randomly generated geometries for the system under scrutiny, whose energy is evaluated at the target level of theory. A subset of these starting candidates is employed to train a GPR MLIP, which is used to evaluate the new candidates generated in subsequent steps and to relax these candidates in the lower confidence bound. From these newly generated candidates, the best one according to an \textit{acquisition} function (which depends on the predicted stability) is selected at the end of each iteration to be evaluated at the target level of theory and incorporated into the dataset for training the MLIP model. Thus, at each step, the accuracy of the model is improved, which in turn improves the selection process for new candidates to the global minimum. Since the energy predictions of the GPR model are based on a similarity measure of the predicted structure with that of the structures employed to train the model, the predictions have an associated uncertainty estimation. This uncertainty measure is incorporated into the acquisition function to bias the search towards exploration of regions of the configurational space which the model is less certain of its predictions. At each iteration, the method generates new candidates either randomly or by modifying previously generated candidates, improving the candidates generated in previous iterations.

Adapting the GOFEE algorithm to work with multiple stoichiometries in a GCGO approach required addressing three main issues. First, the acquisition function of GOFEE depends on the potential energy of the system instead of the Gibbs energy, which is required to compare the stability of systems with different compositions. Furthermore, while the Valle-Oganov fingerprint \textit{descriptor} used in GOFEE’s GPR model produces useful descriptions when training single-stoichiometry models, its stoichiometry-dependent normalization factor makes the descriptor unsuitable for training size-extensive models. We note that by size-extensibility we refer to the correct scaling of the potential energy of the system with the number of atoms. This is not directly related to the \textit{ab initio} thermodynamics framework, although size extensive MLIP are obviously necessary to compute reliable relative Gibbs energies. Finally, new operations were required that modify the composition of the candidates and enable the exploration of different stoichiometries. 

In order to evaluate the Gibbs energies of the candidates during the GCGO executions, we implemented a new \textit{acquisitor} module and two new \textit{sampler} modules into the AGOX library. Sampler modules are responsible for the selection of previously generated candidates from the database in order to seed the candidate generator modules, which create candidates by modifying these seed structures. The new samplers are the \textbf{Gibbs energy of formation sampler} and the \textbf{K-Means Gibbs energy of formation sampler}. Both modules select the seed structures based on their Gibbs energy of formation. The new acquisition function follows the \textit{ab initio Thermodynamics} framework, where Gibbs energy of formation for a given candidate $\Delta G_c$ can be expressed as a function of the chemical potential of reservoirs assumed to be in equilibrium with the system:

\begin{equation}
    \Delta G_c = G_c - G_r - \sum_{v\in V}\Delta n_v\mu_v
\end{equation}

where $G_c$ and $G_r$ are the Gibbs energy of the candidate and a reference system, respectively, $\Delta n_v$ is the difference in number of atoms of every element or adsorbate $v$ between the candidate and the reference system, and $\mu_v$ is the chemical potential of every species $v$ in their reservoir. For a fast evaluation of $\Delta G_c$ of condensed matter, it is common and a valid approximation\cite{thermo-Reuter2003,thermo-Reuter2005,goga-Reichenbach2019}, to neglect the vibrational contributions to $G_c$ and $G_r$, leading to  

\begin{equation}
    \Delta G_c = E_c - E_r - \sum_{v\in V}\Delta n_v( E_v+\Delta\mu_v)
\end{equation}

where $E_c$, $E_r$, and $E_v$ are the potential energy of the candidate, of the reference system, and of each reservoir molecule at 0 K, respectively, which can be provided by the targeted level of \textit{ab initio} theory or the MLIP potential. The $\Delta\mu_v$ terms thus account for temperature and pressure-dependent parts of the chemical potential of the reservoirs.

For the M$_3$O$_x$ test systems evaluated below, we express all $\Delta G_{M_3O_x}$ with respect to a common reference M$_3$, as
\begin{equation}
\Delta G_{M_3O_x}= G_{\text{M}_3\text{O}_{x}} - G_{\text{M}_3} - x \mu_{\mathrm{O}}.
\end{equation}

Since we are considering systems in equilibrium with a O$_2$ gas reservoir,

\begin{equation}
\mu_{\mathrm{O}} = \frac{1}{2} \mu_{\mathrm{O}_2}(T, p) = \frac{1}{2} [E_{\mathrm{O}_2}^{DFT}+ \Delta\mu_{\mathrm{O}_2}(T, p)],
\end{equation}

where $E_{\mathrm{O}_2}^{DFT}$ is the zero-point energy corrected DFT energy and $\Delta\mu_{\mathrm{O}_2}(T, p)$ includes the $T$ and $p_{O_2}$ dependent terms, i.e.,

\begin{equation}
\Delta\mu_{\mathrm{O}_2}(T, p) =  k_B T \ln\left(\frac{p}{p^\circ}\right) + \Delta U_{\mathrm{O}_2}^{\text{thermal}} - T S_{\mathrm{O}_2},
\end{equation}

where $k_B$ is the Boltzmann constant, $\Delta U_{\mathrm{O}_2}^{\text{thermal}}$ are the thermal contributions to the internal energy and $S_{\mathrm{O}_2}$ is the entropy of O$_2$. The $\Delta U_{\mathrm{O}_2}^{\text{thermal}}$ and $S_{\mathrm{O}_2}$ terms can be interpolated from thermodynamic tables or directly calculated by means of frequency calculations within the harmonic oscillator approximation.

As stated above, we neglect thermal contributions to $G_{\text{M}_3\text{O}_{x}}$ and $G_{\text{M}_3}$ and directly approximate these to $E_{\text{M}_3\text{O}_{x}}^{\text{DFT}}$ and $E_{\text{M}_3}^{\text{DFT}}$. This choice is motivated by the much higher cost of calculating the frequencies or thermal contributions for every candidate structure and the resemblance between phase diagrams with and without this simplification. Thus, the algorithm evaluates the Gibbs energy of each candidate only by single-point, zero-Kelvin, potential energy evaluations instead of dynamic simulations or other more elaborate methods. It is therefore by no means a formal operando approach, but rather an efficient algorithm to sample stable structures and stoichiometries. 

To illustrate the effect of vibrational contributions to predicted phase diagrams, we have calculated, for Ir$_3$O$_x$ and directly with DFT calculations, the $p$,$T$ phase diagrams both neglecting and including thermal contributions to $G_{\text{M}_3\text{O}_{x}}$ (see Figure S1 in the SI). 
These are obtained by calculating the vibrational frequencies of each global minimum using the harmonic approximation, which provides the zero-point energy corrections and the vibrational partition function. In both diagrams in Figure S1, vibrational contributions to $\Delta\mu_{\mathrm{O}_2}(T, p)$ are properly accoutned for. This comparison indeed reveals a good qualitative agreement between the two phase diagrams, especially for moderate temperature and pressure ranges. In addition, despite the differences in the location of the phase boundaries, the same main oxidation states appear as stable in the diagram. For further refinement of the results provided by our GCGO algorithm, however, it would indeed be recommendable to carry out the vibrational frequency analysis of the most stable states identified during the search.

To run the GCGO algorithm on the selected target T and $p_{\mathrm{O}_2}$, one can then simply evaluate the resulting (constant) $\Delta \mu_{O_2}(T, p_{\mathrm{O}_2})$ and calculate $\Delta G_c$ of every candidate i as

\begin{equation}
    \Delta G_{i, M_3O_x} = E_{i, M_3O_x} - E_{r,M_3} - \frac{x}{2} [E_{\mathrm{O}_2}^{DFT}+ \Delta\mu_{\mathrm{O}_2}(T, p_{O_2})].
\end{equation}

The \textbf{Gibbs energy of formation sampler} selects the $N$ previously generated candidates with the lowest Gibbs energy of formation to seed the generator. In turn, the \textbf{K-Means Gibbs energy of formation sampler} performs a \textit{K-means} clustering of all the previously generated candidates, categorizing them by the similarity of the sum of the SOAP atomic center description into a selected number of clusters and choosing the most stable candidate of each cluster.

Once the generators have created new candidates, the \textbf{Gibbs energy of formation acquisitor} evaluates their Gibbs energy with the MLIP and selects the fittest candidate(s) to be evaluated by the target level of theory (and included in training set of the MLIP). The uncertainty measure is incorporated into the acquisition function as
\begin{equation} \label{kappa}
    F = \Delta G^{M}_c - \kappa \sigma
\end{equation}
where $\Delta G^{M}_c$ is the Gibbs energy of formation of the candidate evaluated with the model, $\kappa$ is a user defined parameter and $\sigma$ is the uncertainty of the prediction by the MLIP. The acquisitor thus selects the candidates with the lowest value of F, favoring the first principles evaluation of candidates that are predicted to be stable by the model and which are dissimilar to previous candidates that the MLIP has been trained on (i.e., unexplored regions of the potential energy surface). 

The GCGO algorithm relies on the MLIP termed \textit{local surrogate model} in the AGOX package\cite{agox-Christiansen2022,agoxlgpr-Rnne2022} and based on the Gaussian Approximation Potential (GAP) approach \cite{gap-Bartk2010,gap-Bartk2017,gap-Chmiela2017,gap-Deringer2018,gap-Klawohn2023}. This approach combines GPR with the local Smooth Overlap of Atomic Positions (SOAP) descriptor\cite{locald-Bartk2013}, which separates each structure or candidate into atomic environments, generating a local description for each atom of the system. During the training of this model, the energies of each structure are decomposed into atomic contributions. Since potential energy predictions of new structures are also obtained as the sum of their atomic contributions, the model is size extensible, i.e. can be used for structures with different stoichiometries (i.e., compositions). However, due to the large number of training points generated (several for each stoichiometry), a sparse GPR is employed, using CUR decomposition to choose the inducing points\cite{cur-Mahoney2009, agoxlgpr-Rnne2022}.

To generate new candidates by modifying the composition of the previously explored structures and enable exploring different stoichiometries, we implemented the \textbf{Addition and Removal Generator}. This generator creates candidates by adding or removing atoms from previously selected candidates. The generator takes a seed structure as input and a randomly chosen composition target (within the allowed range). It then adds and/or removes atoms from the seed structure until the target composition is achieved. This generator thus allows obtaining candidates of different stoichiometries without randomly generating them, preserving stable motifs across compositions.

In addition to these new modules, other modifications of the GOFEE approach in AGOX include the \textbf{Environment} module to work with multiple, user-selected stoichiometries, and other utilities.


\subsection{Hyperparameters}

The parameters of the GCGO algorithm are similar to those of the original single-stoichiometry GOFEE code, which we have adapted heuristically for the performance tests:

\begin{itemize}
\item $\kappa$: as indicated in eq. \mbox{\ref{kappa}}, this parameter controls the weight given in the acquisition function to the uncertainty of the surrogate energy model. The GOFEE default is typically 1 for single stoichiometries, which we double for the GCGO runs to further favor exploration of unvisited regions of the configurational and compositional spaces.
\item Number of candidates generated at each step: we set the value to 32 to best adapt to the 32-processor nodes used for running the GCGO algorithm, which allows us to parallelize candidate generation within our computing architecture. Note that of these 32 candidates, 8, 12, and 12 are generated with the Random, Rattle, and Addition/Removal generators, respectively. These ratios are different during the initialization - for the first five steps all candidates are randomly generated, and for the next five steps 8 candidates are generated with the Random generator and 24 with the Rattle generator.
\item Number of previously calculated structures that are employed by the sampler to create new candidates: since the default is 10 for single stoichiometry GOFEE runs, we set 20 for GCGO. This is because the KMeans sampler creates a number of clusters equal to the number of samples, and we wanted to ensure that clusters do not only separate by stoichiometry.
\end{itemize}

\subsection{Test systems}

It is important to note that our GCGO algorithm is agnostic to the level of theory (or software) used to compute the energy of the target system and can therefore be used to search for stable chemical states and structures of target systems whose potential energy surfaces are described at any chosen level of theory, independent of how accurately the chosen level of theory represents real scenarios. Nevertheless, the computational benefit of our GCGO algorithm is obviously most prominent in applications where the MLIP trained on-the-fly is used as a surrogate energy model of first principles levels of theory (e.g. DFT). We illustrate the performance of the implemented code by applying it to three different test systems with different target levels of theory. Namely, we evaluate the capacity of the GCGO algorithm to identify the most stable structure and chemical state for free-standing Ir$_3$O$_x$ clusters, Pt$_3$O$_x$ clusters supported on the CeO$_2$(111) surface, and oxidized states of the Pd(100) surface at conditions with various degrees of oxidizing or reducing character (i.e., higher or lower chemical potentials $\mu_{O_{2}}$ of an O$_2$ reservoir). For the Ir$_3$O$_x$ clusters, the target level of theory is a MACE\cite{mace-Batatia2022Design} MLIP trained on Ir$_y$O$_x$ clusters. Using this affordable target level of theory allows running the GCGO algorithm numerous times but starting from different initial random seeds to get statistically significant data about the cumulative success rates. For the optimization of Pt$_3$O$_x$ clusters supported on the CeO$_2$(111) surface and for the Pd(100) surface, we employ DFT calculations as the target level of theory (see details below), showing the applicability of the approach when first principles calculations are required.

\begin{figure}
\centering
\includegraphics[width=0.5\textwidth]{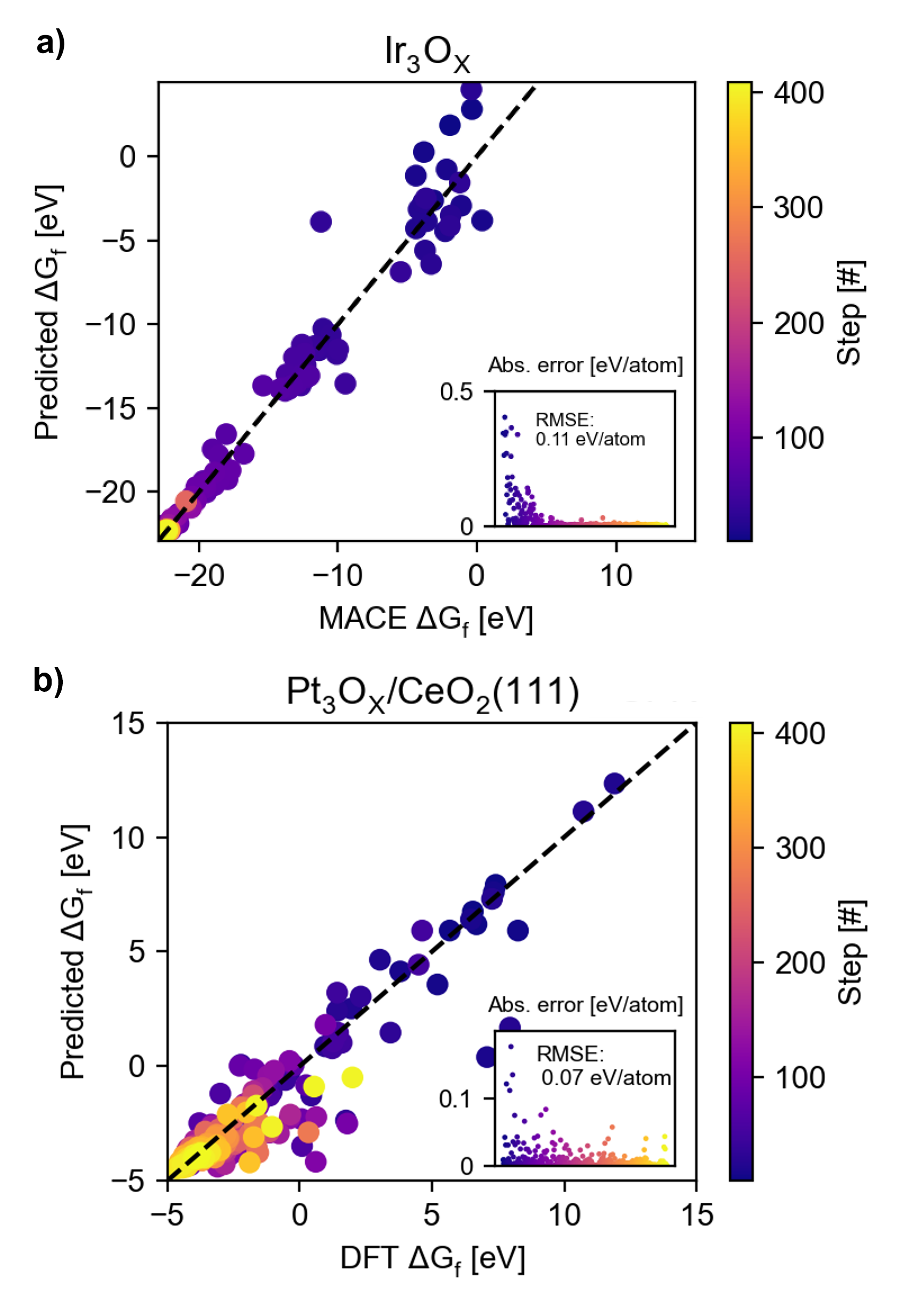}
\caption{\textbf{On-the-fly learning performance of the SOAP+GPR MLIP for Gibbs energies}. Comparison of predicted and reference energies (parity plots) during single executions of the GCGO algorithm for optimizing the structure and chemical state of \textbf{a)} free standing Ir$_3$O$_x$ particles and \textbf{b)} Pt$_3$O$_3$ particles supported on CeO$_2$(111). Color of the marker indicates step number according to the color-bar reference. Insets show the corresponding absolute error as a function of step number.}
\label{parity}
\end{figure}

For all systems, we employ the \textbf{K-Means Gibbs energy of formation sampler} with 10 samples on each step. 32 new candidates are generated in each step, with the Rattle and \textbf{Addition and Removal} generators starting to create new candidates from iterations 5 and 10 respectively.
We employ a $\kappa=2$ value for the \textbf{Gibbs energy of formation acquisitor}, this acquisitor is employed in both the energy evaluations and the lower confidence bound relaxations.
The SOAP descriptor parameters used are $R_{cut}=5 $, $n_{max}=3$, $l_{max}=2$, $\sigma=1$. The GPR model employs a \textit{Radial Basis Function} kernel and the CUR algorithm by Mahoney and Drineas \cite{cur-Mahoney2009} to select up to 1000 inducing points for the sparsification procedure.
The resulting model performs well for on-the-fly training during the GCGO code executions, as illustrated in Fig. \ref{parity}. The evolution of the Gibbs formation energy ($\Delta$G$_f$) of evaluated candidates along a single GCGO run is shown for both free standing Ir$_3$O$_x$ particles (Fig. \ref{parity}a) and Pt$_3$O$_3$ particles supported on CeO$_2$(111) (Fig. \ref{parity}b).
The insets show the evolution of the absolute error of the SOAP+GPR MLIP trained on the fly, calculated at each step N for the SOAP+GPR MLIP trained with the energies of the previously evaluated N-1 candidates.

These runs evolve as expected, with both the $\Delta$$G_f$ values of the explored candidates and the absolute errors of the surrogate energy model getting progressively lower along the run. The GPR model thus offers reasonable accuracy with RMSE values of 0.11 and 0.07 eV/atom for the Ir$_3$O$_x$ and Pt$_3$O$_3$/CeO$_2$(111) runs, respectively. However, we must note that these RMSE values include the initial prediction errors that are, as expected, quite large as the model is then trained only with small training sets. The RMSE calculated only for the last 100 steps of the Pt$_3$O$_3$/CeO$_2$(111) in Fig. \ref{parity}{b is of just 8 meV/atom, indicating that, specially at more advanced stages of a run, the surrogate energy model is accurate enough to evaluate the stability  of candidate structures with reasonable fidelity.

To further evaluate the performance of the GPR+SOAP model, we have compared its accuracy and training cost to that offered by more modern MLIP architectures, namely NequIP \cite{Batzner2022NequIP} and MACE \cite{mace-Batatia2022mace} (See Section S2 in the Supplementary information file). Training cost is a very relevant metric for on-the-fly trained potentials because, for larger models, it can represent a bottleneck in the algorithm. As shown in Figure S2, the MACE and NequIP training costs make these potentials a much less convenient option than the GPR+SOAP model for on-the-fly training. The corresponding parity plots and RMSE values indicate that these larger training costs do not lead to more accurate MLIPs within the low-data-size regimes in typical GCGO runs (Figure S2 and S3), indicating that the GPR+SOAP model is both more accurate and computationally efficient.

Another very relevant advantage of GPR-type models is their direct evaluation of the prediction uncertainty at no extra computational cost. Because the prediction uncertainty is inferred directly from the similarity of a candidate structure to those in the training set, this uncertainty can be straightforwardly included in the acquisition function of our algorithm to promote exploration of structures distinct from those already evaluated by DFT. At present, the most well-established approach for estimating prediction uncertainty in models such as MACE or NequIP is the ensemble (or committee) method, whereby multiple instances of the same model are trained, and the uncertainty is quantified by the standard deviation of their predictions} \cite{farris2025bayesianneuralnetworksversus}. Training various instances of models such as MACE or NequIP to estimate uncertainty would further exacerbate the already high training costs of these models. In addition, in GPR, the uncertainty estimations in neural networks are a consequence of the loss function reaching different local minima, which is, at best, only indirectly related to structural dissimilarity.

\subsubsection{Grand canonical global optimization for free-standing Ir$_3$O$_x$ particles}\label{perf-section}

For the free-standing Ir$_3$O$_x$, we carry out a statistical analysis of the performance of the GCGO algorithm in the form of calculated cumulative success rates. A run of the algorithm is considered successful if the Gibbs Energy of formation of the most stable configuration is less than $0.5$ eV higher than the true most stable candidate found at the considered environmental conditions. This criterion (for systems of $\approx$ 10 atoms) resembles the relatively arbitrary choice of 0.05 eV/atom used in previous evaluations of the Canonical (fixed stoichiometry) algorithms within AGOX (see \mbox{\cite{agoxlgpr-Rnne2022, agox-Christiansen2022, gofee-Bisbo2020, gofee-Bisbo2022}}) and allows us to compare cumulative success rates of grand canonical runs versus those of single stoichiometry searches. We also note that using a per atom criterion for $\Delta$G$_f$ would lead to structures with more atoms being more likely to pass the success criterion than equally stable structures with fewer atoms.

This analysis thus relies on a known set of global minima and resulting phase diagrams for the Ir$_3$O$_x$ system. Cumulative success rates are then calculated as a function of the number of iterations (i.e., number of single-point energy evaluations with the target level of theory) as the percentage of the code executions that have been successful at each iteration step. We thus run 25 code executions for each of the three distinct environmental conditions, namely $\Delta\mu_O = 0.0$ eV, $\Delta\mu_O = -1.15$ eV, and $\Delta\mu_O = -1.6$ eV. For each execution, the code is allowed to run for 1000 steps (1000 single-point evaluations). 

Given the large number of code executions required to obtain statistically significant data, we use a MLIP based on the MACE architecture\cite{mace-Batatia2022Design,mace-Batatia2022mace} as the target level of theory. This MACE model was trained on a database of Ir$_3$O$_x$ structures and their energies obtained by means of single point Density Functional Theory (DFT) calculations using the Fritz Haber Institute ab initio materials simulations (FHI-AIMS)\cite{fhi-aims} with PBE exchange correlation functional\cite{pbe-Perdew1996} and an ultralight-tier 1 numerical atom-centered orbital (NAO) basis set with a relativistic zeroth order approximation.

\subsubsection{Grand canonical global optimization for Pt$_3$O$_x$ particles on CeO$_2$(111)}

In order to demonstrate the capacity of the GCGO code to identify stable structures and chemical states when using first principles methods as the target level of theory, we reproduce the results obtained in a previous work employing the GOFEE method for the catalytically relevant Pt$_3$O$_x$/CeO$_2$(111) system\cite{pt3ox-QuinlivanDomnguez2022}. Stable structures for the Pt$_3$O$_x$/CeO$_2$ system were found by using the GOFEE method for each of the $x\in\{1,...,6\}$ considered stoichiometries. This involved 800 single-point evaluations per considered stoichiometry at the target DFT level of theory, using the semi-local PW91 exchange-correlation functional\cite{pw91-Burke1998,pw91-Perdew1992} as implemented in the VASP code\cite{vasp-Kresse1993,vasp-Kresse1996,vasp2-Kresse1996,vasp-pp-Kresse1994}. The remaining computational details for these DFT calculations are the same as those used in \cite{pt3ox-QuinlivanDomnguez2022} for the single-point evaluations when running GOFEE. 

We again consider three different environmental conditions corresponding to $\Delta\mu_O$ values of 0.0, -0.5, and -1.0 eV and perform 10 GCGO searches for the Pt$_3$O$_x$/CeO$_2$(111) with $x\in\{1,...,6\}$ systems. The CeO$_2$ surface is represented by a single O-Ce-O trilayer. Each independent GCGO execution consists of 1500 iterations, and on each iteration two single point evaluations are performed, one of the selected candidate and a second evaluation after advancing the candidate a single structural relaxation step, adding up to a total of 3000 single point evaluations per execution. The remaining run parameters are identical to the ones employed for \textbf{Section \ref{perf-section}}. Performance is evaluated after 400 and 1500 iterations (800 and 3000 single-point DFT calculations, respectively).

\subsubsection{Grand canonical global optimization for the Pd(100) surface}
To show that the GCGO algorithm is able to find reconstructions not only of clusters, but also of extended systems, we tackle the Pd(100) surface and its known stable phases under oxidative conditions. As many late transition metal surfaces, the chemical state of Pd(100) depends on the conditions it is exposed to, resulting in either pristine surfaces, oxygen-decorated states that preserve the surface structure, or surface-oxide phases where the outermost layer is strongly reconstructed \cite{Pd100-Rogal2007}. 

We use the GCGO algorithm to investigate this system, by performing 15 searches at each of five different chemical potentials for oxygen, $\Delta \mu_O$ = [-3.0, -2.5, -2.0, -1.5, -1.0], adding up to a total of 45 GCGO runs.
Each search consists of 500 iterations (single-point DFT evaluations) and is allowed to explore stoichiometries 
with $N_{Pd} = [4, 5]$ and $N_O = [0, 1, 2, 3, 4]$. The searches are done in the 
$\sqrt{5} \times \sqrt{5}R27^\circ$ cell that is able to accommodate the surface-oxide reconstruction.
This choice of cell allows for the correct description of the bare surface and the 
reconstruction. It is, however, not the optimal cell for the most stable (2$\times$2) phase with adsorbed oxygen. However,
a phase with adsorbed oxygen is found to be preferential for a range 
of chemical potentials, so this choice of cell leads to qualitatively correct 
results. The system is evaluated using DFT, as implemented in GPAW\cite{gpaw-Mortensen2024}, using the PBE functional\cite{pbe-Perdew1996} and a 2$\times$2 Monkhorst-Pack grid for k-points and 
a cutoff of 400 eV for the plane-wave basis set.

\section{Results}

\subsection{Performance analysis for free-standing Ir$_3$O$_x$ particles}\label{sec1}
\begin{figure}
\centering
\includegraphics[width=0.45\textwidth]{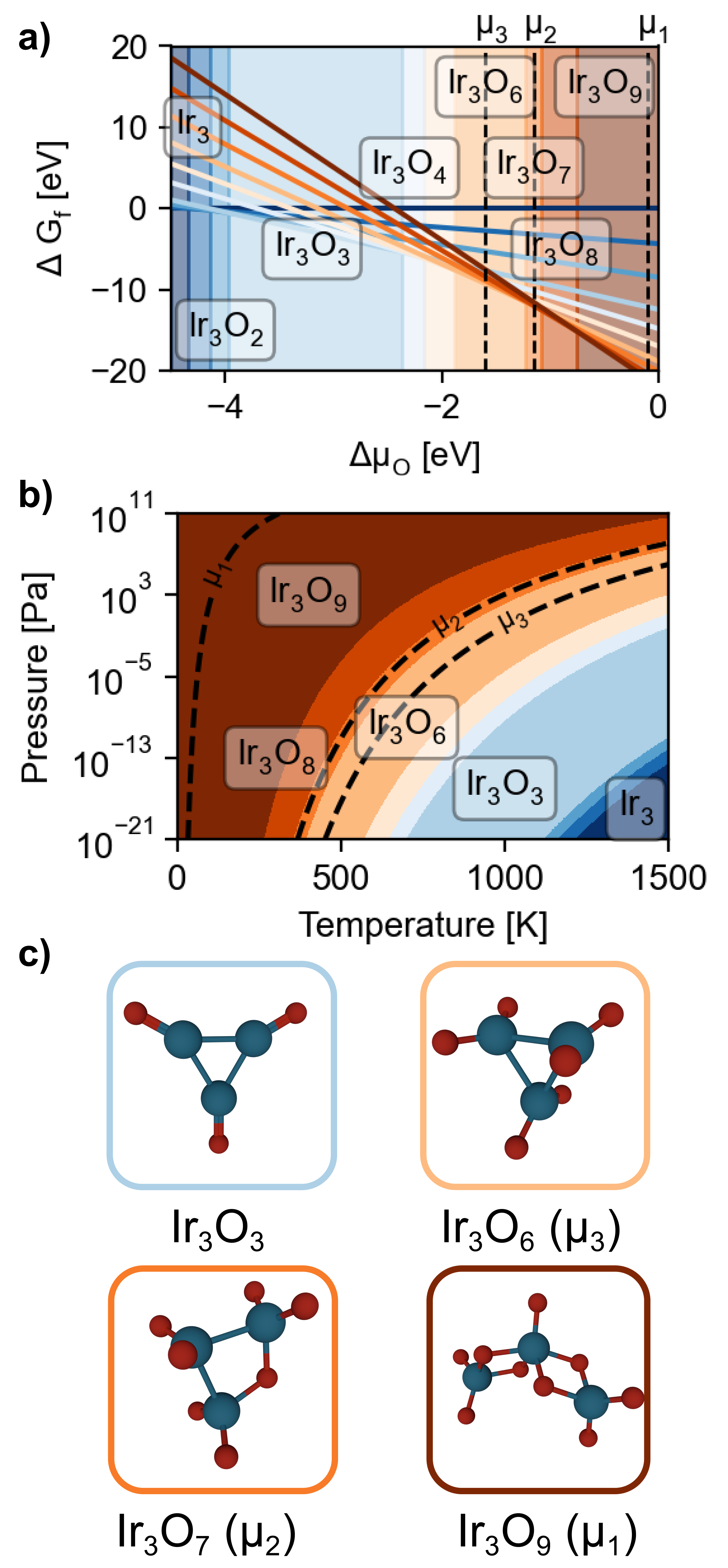}
\caption{\textbf{Ir$_3$O$_x$ phase diagram}. \textbf{a)} $\Delta G_f$ vs $\Delta\mu_O$ phase diagram for the Ir$_3$O$_x$ system at the target MACE level of theory constructed with the best obtained candidates for each stoichiometry. Every line of corresponds to the evolution of $\Delta G_f$ of the best structure of each stoichiometry as a function of $\Delta\mu_O$, and colors indicate, for each region, which of the considered stoichiometries has the lowest $\Delta G_f$ (i.e., is the most stable). Dotted lines indicate the $\Delta\mu_O$ values that have been targeted with the GCGO algorithm: $\mu_1=$ 0.0 eV, $\mu_2=$ -1.15 eV, $\mu_3 =$ -1.6 eV. \textbf{b)} $p_{O_{2}}$,$T$ phase diagram for the same system, constructed by evaluating $\Delta\mu_O$( $p_{O_{2}}$,$T$) at different $p_{O_{2}}$ and $T$ values. \textbf{c)} Structures of the most stable Ir$_3$O$_x$ state for different $\Delta\mu_O$ values.}
\label{irox-search1}
\end{figure}

We start by demonstrating that the GCGO algorithm is capable of efficiently identifying stable structures and chemical states of targeted systems and conditions. In each execution of the GCGO code for the Ir$_3$O$_x$ system, stoichiometries with $x\in\{0,...,9\}$ are explored at three different environmental conditions defined by the corresponding oxygen chemical potential $\Delta\mu_O$. We sample $\Delta\mu_O$ values of 0.0 ($\mu_1$), -1.15 ($\mu_2$), or -1.6 eV ($\mu_3$), corresponding to very oxidizing, mildly oxidizing, and reducing regions, respectively, of the phase diagram shown in Fig. \ref{irox-search1}a. In order to evaluate the performance with statistical significance, 25 executions are carried out for each $\mu$.

Fig.\ref{irox-search1}a shows the $\Delta G_f$ vs $\Delta\mu_O$ phase diagram of Ir$_3$O$_x$ at the MACE MLIP target level of theory, with each individual line corresponding to the evolution of $\Delta G_f$ of the global minimum of each stoichiometry.
Fig. \mbox{\ref{irox-search1}}b corresponds to the same phase diagram, but evaluating $\Delta\mu_O$($p_{O_{2}}$,$T$) at different  $p_{O_{2}}$ and $T$ values. The most stable state (that with lowest $\Delta G_f$) for each region is indicated in both diagrams and the corresponding structures are illustrated in Fig.\mbox{\ref{irox-search1}}c. The global minima correspond to the best structures found for each stoichiometry along the 75 GCGO runs as well as in single-stoichiometry runs of the GCGO algorithm carried out as consistency tests. The $\Delta\mu_O$ values ($\mu_1$, $\mu_2$, and $\mu_3$) that have been targeted during the GCGO runs are shown as dotted lines in Fig.\mbox{\ref{irox-search1}}a and b.

The cumulative success rate curves in Fig. \mbox{\ref{irox-search1}}a (solid lines) correspond to the percentage of the 25 GCGO executions per targeted $\Delta\mu_O$ that has found the most stable structure and chemical state at that target $\Delta\mu_O$ a. That is, the proportion of runs that have found the global minimum illustrated in Fig. \mbox{\ref{irox-search1}}c  (or one up to 0.5 eV higher $\Delta G_f$) as a function of the number of executed steps (single point evaluations at the target level of theory). These success rates are compared in Fig. \mbox{\ref{irox-search1}}a to those of single-stoichiometry GO runs for the stoichiometries of the most stable structures at $\mu_1$, $\mu_2$, and $\mu_3$ (i.e., Ir$_3$O$_9$, Ir$_3$O$_7$, and Ir$_3$O$_6$, respectively). The comparison these single-stoichiometry runs allows us to establish the computational efficiency of the GCGO algorithm with respect to to the more common fixed-stoichiometry approach and to rationalize why a GCGO is less successful for searches at given target $\Delta\mu_O$.

\begin{figure}\
\centering
\includegraphics[width=0.45\textwidth]
{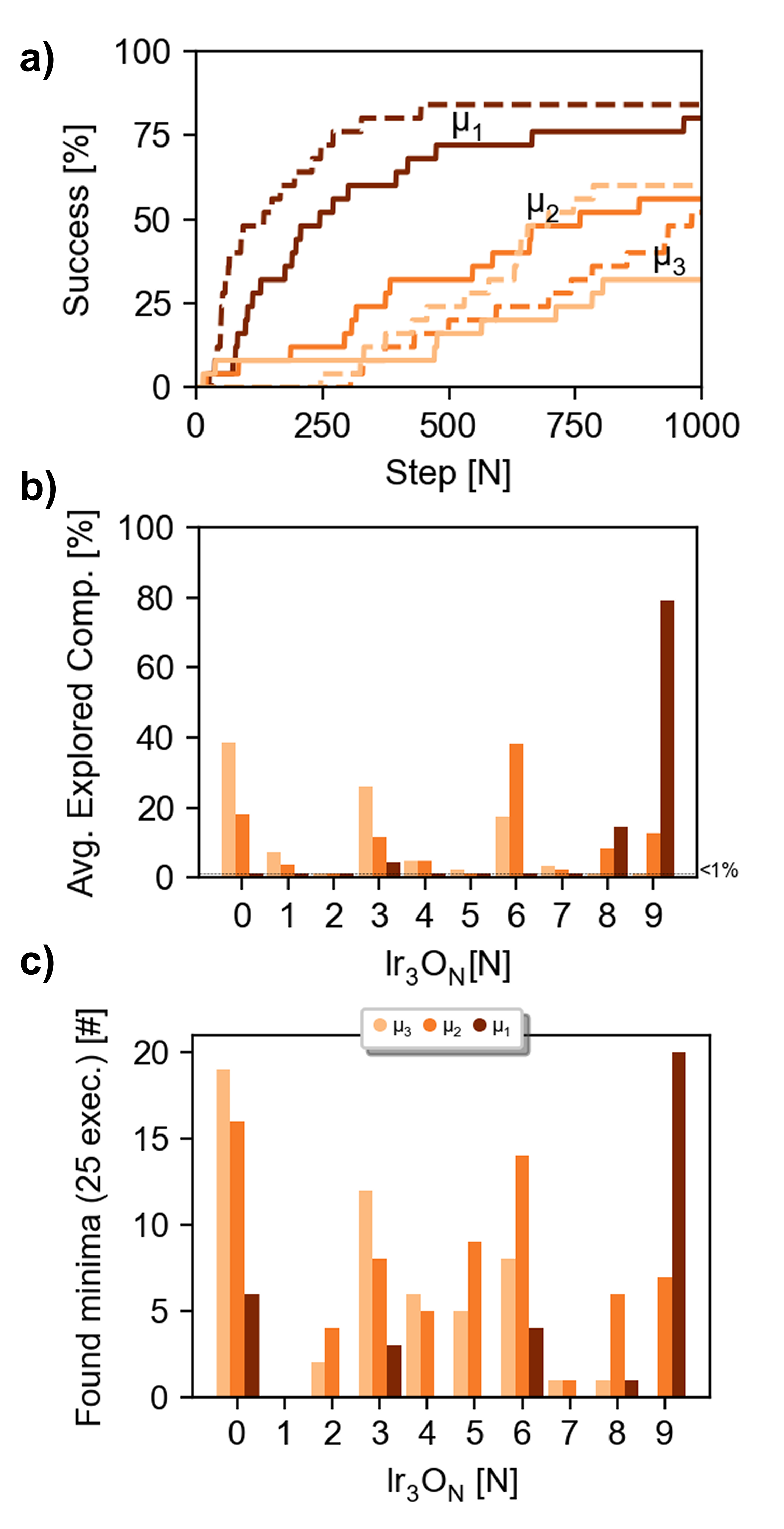}
\caption{\textbf{a)} Cumulative success rates of the GCGO runs (solid lines) for free-standing Ir$_3$O$_x$ at $\mu_1$, $\mu_2$, and $\mu_3$. Dotted lines correspond to the single-stoichiometry GO runs for the Ir$_3$O$_x$ stoichiometries of the most stable structures at $\mu_1$, $\mu_2$, and $\mu_3$ (i.e., Ir$_3$O$_9$, Ir$_3$O$_7$, and Ir$_3$O$_6$, respectively). \textbf{b)} Distribution of evaluations per stoichiometry across the 25 GCGO for each target potential. \textbf{c)} Number of GCGO optimizations (out of 25 runs) that have found the global minimum of different stoichiometries as a function of the targeted $\mu$.}
\label{irox-search2}
\end{figure}

The calculated success rates indicate that the GCGO algorithm can indeed efficiently identify stable structures at different target $\Delta\mu_O$, although the performance depends strongly on the targeted $\Delta\mu_O$. This points to an effect of the complexity of the evaluated region of the phase-diagram and of the conformational space of the stoichiometries that are stable at each potential. The distribution of explored stoichiometries in Fig. \mbox{\ref{irox-search2}}b also demonstrates the capacity of the GCGO to explore various oxidation states, focusing on stoichiometries around that of the global minimum.

Remarkably, the success rates of GCGO runs exploring various stoichiometries are comparable to those of single-stoichiometry GO runs. In the case of GCGO runs at the target potentials $\mu_1$ and $\mu_3$, the success rates are only slightly lower than in the single-stoichiometry searches of the global minimum for the corresponding stoichiometries (Ir$_3$O$_9$ and Ir$_3$O$_6$, respectively). For the GCGO runs at $\mu_2$, surprisingly, the success rate of the GCGO is even higher than that of single-stoichiometry searches for Ir$_3$O$_7$. This, however, is related to the choice of the success criterion, which is defined as finding a structure within 0.5 eV (in terms of Gibbs energy) of the global minimum, independently of the stoichiometry of that structure. The $\mu_2$ target potential corresponds to a narrow region of the phase diagram where the global minimum of Ir$_3$O$_7$ is the most stable stoichiometry, but where the global minima of the Ir$_3$O$_6$ and Ir$_3$O$_8$ are close in energy. GCGO runs at this $\mu_O$ target potential are therefore sometimes considered successful also when a stable enough structure of Ir$_3$O$_6$ or Ir$_3$O$_8$ is found. This drives the success rate of the GCGO up compared to the single-stoichiometry Ir$_3$O$_7$ runs, which can only sample the presumably more complex potential energy surface of Ir$_3$O$_7$. This is corroborated by the distributions of single-point evaluations for GCGO runs at $\mu_2$= -1.15 eV (\mbox{\ref{irox-search2}}b), which show that these runs have indeed focused more on Ir$_3$O$_6$ and Ir$_3$O$_8$ structures than Ir$_3$O$_7$. Adding the requirement that for a run to be successful that it correctly identifies the stoichiometry of the real global minimum (Ir$_3$O$_7$ for $\mu_2$) lowers the success rate to ($\sim10\%$). Obviously, such lower success rates are to be expected when only solutions with the right stoichiometry are considered valid, especially when targeting $\Delta\mu_O$ values corresponding to narrow regions of the phase diagram.

Inferring the complexity of the conformational space for each stoichiometry from the success rates of single-stoichiometry GO searches shows that, counterintuitively, there is not a clear correlation between the size of the system and the resulting complexity. For the Ir$_3$O$_x$ system, the global minimum of Ir$_3$O$_9$ is the easiest to find, which explains why GCGO runs at $\mu_3$, for which Ir$_3$O$_9$ is the most stable stoichiometry, have the highest success rates. Single-stoichiometry GO runs for Ir$_3$O$_6$ and Ir$_3$O$_7$ have similarly lower success rates, which contributes to GCGO success rates at $\mu_2$ and $\mu_3$ being also similar. For GCGO runs at these $\mu_O$ values, the distributions per stoichiometry of energy evaluations along a run (Fig.\mbox{\ref{irox-search2}}b)seem biased to sample structures with lower O contents than the global minima. This could be due to smaller systems requiring fewer steps to reach stable configurations, biasing searches towards local minima of states with fewer atoms or directly to the lower complexity of the conformational space of some stoichiometries. Note that in the case of GCGO runs at $\mu_3$, a significant fraction of the evaluated structures have Ir$_3$ and Ir$_3$O$_3$ stoichiometry instead of the most stable Ir$_3$O$_6$. Thus, despite these stoichiometries becoming stable only at much lower $\Delta\mu_O$ values, the  GCGO search seems biased towards exploration of these less reduced states.

The exploration of various stoichiometries along a single run results in one of the most notable strengths of the GCGO algorithm: a single run can identify the global minima of various stoichiometries. This is illustrated in Fig.\mbox{\ref{irox-search2}}c, where we collect the number of runs, as a function of the targeted $\Delta\mu_O$, that have found the minimum energy structures of each stoichiometry. Despite targeting $\Delta\mu_O$ values for which the global minimum corresponds to either Ir$_3$O$_9$, Ir$_3$O$_7$, or Ir$_3$O$_6$, runs at these potentials regularly find the most stable structures of Ir$_3$, Ir$_3$O$_2$, Ir$_3$O$_3$, Ir$_3$O$_4$, Ir$_3$O$_5$, or Ir$_3$O$_8$. 

Summarizing this performance analysis, the GCGO algorithm is capable of minimizing the Gibbs energy of formation for a given target $\Delta\mu_O$, thus finding candidates with stable structures and compositions for the targeted environmental conditions. Remarkably, the algorithm is capable of finding the global minimum at a given target $\mu_O$ with a similar computational cost as single-stoichiometry runs while, in addition, finding the most stable structures of various other stoichiometries.

\subsection{Performance analysis for Pt$_3$O$_x$ particles on CeO$_2$(111)}

After having demonstrated the performance of the implemented GCGO code on free clusters described by an MLIP, we illustrate its effectiveness in finding the global minimum of a catalytically relevant target material at the DFT level of theory. In particular, we reproduce results previously obtained \cite{pt3ox-QuinlivanDomnguez2022} with single-stoichiometry optimizations with the GOFEE algorithm for Pt$_3$O$_x$ clusters supported on the CeO$_2$(111) surface, which can exhibit stoichiometries of $x\in\{0,...,6\}$. 

10 executions of the GCGO were run for each of the three targeted environments, as defined by the $\Delta\mu_0$ values of 0.0, -0.5, and -1.0 eV. Every execution carried out 3000 single-point DFT evaluations, which is comparable to the computational cost incurred in Ref [\citenum{pt3ox-QuinlivanDomnguez2022}] for performing 2-3 executions (per stoichiometry) of the single-stoichiometry GOFEE algorithm with 800 single-point evaluations per execution. We also evaluate the performance after shorter GCGO executions involving 800 single-point DFT evaluations.

\begin{figure}[ht]
\centering
\includegraphics[width=0.40\textwidth]{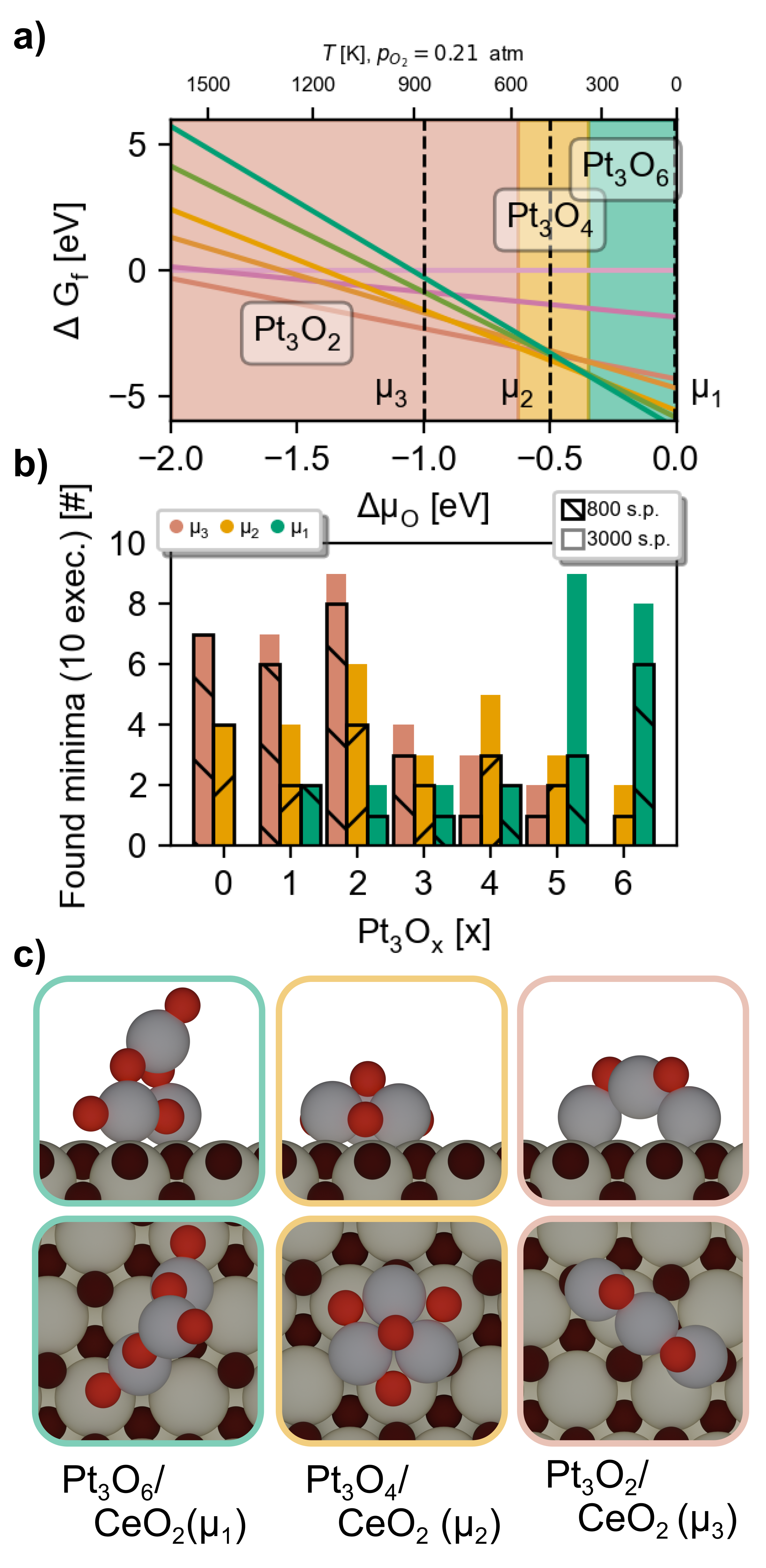}
\caption{\textbf{Performance of the GCGO algorithm for optimization of CeO$_2$(111)-supported Pt$_3$O$_x$ clusters at the DFT level of theory}. \textbf{a)} Phase diagram for Pt$_3$O$_x$/CeO$_2$(111) with $x\in\{0,...,6\}$ at the DFT level of theory. Dotted lines show the $\Delta\mu_O$ chosen for GCGO executions. In the upper axis, $\Delta\mu_O$ is evaluated in terms of temperature at constant pressure. \textbf{b)} Number of GCGO runs, as a function of target $\Delta\mu_O$ and stoichiometry, where a candidate $0.1$ eV less stable or better than the GOFEE-obtained minima was found. Note that a single GCGO execution can identify the minima for several stoichiometries. Hatched and clear bars indicate values obtained after each GCGO was run for 800 and 3000 single-point DFT energy evaluations, respectively. \textbf{c)} Most stable oxidation states and structures identified for each targeted $\Delta\mu_O$.}
\label{pt3ox-search1}
\end{figure}

The phase diagram of the  Pt$_3$O$_x$/CeO$_2$ $x\in\{0,...,6\}$ system is shown in Fig. \ref{pt3ox-search1}a for a $\Delta\mu_O$ range from $-2.0$ to $0.0$ eV, where the three targeted $\Delta\mu_O$ values are shown as a dotted lines. Note that this phase-diagram is different than the one illustrated in Ref [\citenum{pt3ox-QuinlivanDomnguez2022}] because it has been constructed with the data from the single-point DFT calculations of the GOFEE runs only, instead of the more precise data obtained after selected candidates were further optimized using tighter convergence criteria and better surface models. As a measure of the GCGO success rate, Fig. \ref{pt3ox-search1}b shows the number of executions, as a function of target $\Delta\mu_O$ and stoichiometry, where a structure $0.1$ eV less stable or better than the GOFEE candidate to the global minimum was found. Hatched bars correspond to data obtained for the first 800 single-point evaluations of every GCGO execution, which constitutes a direct comparison to the single-stoichiometry GOFEE runs. Plain bars correspond to data from full executions, i.e., with $\sim$ 3000 single-point DFT calculations each. Fig. \ref{pt3ox-search1}c illustrates the structure of the most stable states at each evaluated $\Delta\mu_O$: Pt$_3$O$_6$ for $\mu_1$ (green), Pt$_3$O$_4$ for $\mu_2$ (yellow), and Pt$_3$O$_2$ for $\mu_3$ (red). Remarkably, the GCGO algorithm has, for some stoichiometries, obtained better global minima than in our previous work carried out using single-stoichiometry GOFEE runs. This is the case for the Pt$_3$O$_6$ and Pt$_3$O$_4$ states, whose new global minima structures shown in Fig. \ref{pt3ox-search1}c are $0.70\ eV$ and $0.56 \ eV$ more stable, respectively, than previously identified minima. This is in part due to the capacity of the GCGO to use stable structures of one stoichiometry to generate, by removal or addition of atoms, reasonable candidates of a similar stoichiometry.

Evaluating the performance for the GCGO runs for $\mu_1$, eight of the long executions (3000 steps) and six of the short executions (800 steps) have found the global minimum structure identified in Ref [\citenum{pt3ox-QuinlivanDomnguez2022}] or a more stable one. Furthermore, in several executions at this target potential, global minimum structures for Pt$_3$O$_5$, Pt$_3$O$_4$, Pt$_3$O$_2$, and Pt$_3$O$_1$ have also been found. These data show that the GCGO algorithm not only can reproduce the single-stoichiometry results with lower computational cost at the targeted potential but can also provide promising structures for various other conditions. 

The GCGO executions for $\mu_2$ indicate that, as was the case for Ir$_3$O$_x$ (\ref{sec1}), correctly identifying the global minima structure and oxidation state is more challenging when there are several low-lying metastable states, with five and three executions finding the correct Pt$_3$O$_4$ global minimum after 3000 and 800 single-point evaluations, respectively. Optimizations at this potential also illustrate the bias towards smaller systems, as many of these executions have found the global minima of Pt$_3$O$_2$, Pt$_3$O$_1$, and Pt$_3$. 

The GCGO runs for $\mu_3$ are the most successful among all targeted $\Delta\mu_O$ values, with up to nine (eight) of the ten runs having found the correct global minimum after 3000 (800) steps. This, again, reflects both the smaller problem space of the Pt$_3$O$_2$ stoichiometry, as well as the significantly lower $\Delta G_f$ of the Pt$_3$O$_2$ global minimum compared to those of other stoichiometries.

\begin{figure}[ht]
\centering
\includegraphics[width=0.5\textwidth]{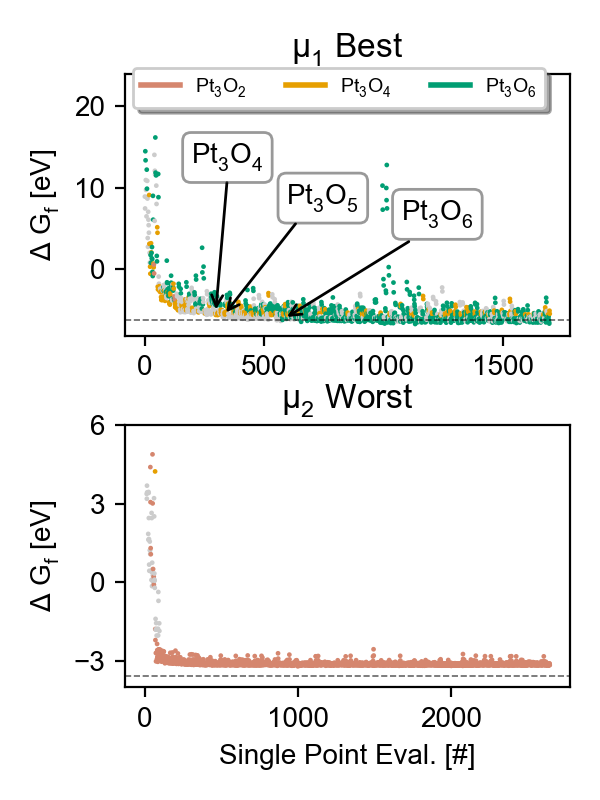}
\caption{\textbf{Examples for successful and unsuccessful GCGO runs for Pt$_3$O$_x$}. Evolution of the Gibbs energy of formation as a function of the number of single-point DFT evaluations during the execution of the run. Colors indicate evaluated stoichiometry at each step, with gray corresponding to stoichiometries not appearing as stable phases in the phase diagram. a) The most successful execution for $\Delta\mu_O=$0.0 eV ($\mu_1$), b) The worst execution at $\Delta\mu_O=$-0.5 eV($\mu_2$). }
\label{pt3ox-search2}
\end{figure}

To further evaluate successful and unsuccessful runs, in Fig. \ref{pt3ox-search2} we show the evolution of the Gibbs energy of evaluated candidates vs the number of GCGO steps (i.e., single-point DFT calculations) for the best run at $\mu_1$ and the worst run at $\mu_2$. Both executions show a general minimization of the Gibbs energy during the execution of the algorithm. The colors indicate the stoichiometry of the evaluated candidate, with gray being used for stoichiometries not appearing as most stable in any region of the phase diagram. The best execution targeting the $\mu_1$ environment shows how the algorithm focuses on the most stable stoichiometry  for $\Delta\mu_O$=0.0 eV (Pt$_3$O$_6$) with occasional explorations into other stoichiometries such as Pt$_3$O$_4$ and Pt$_3$O$_5$, for which it finds new global minima at the steps indicated with arrows. Thus, during this run, the algorithm has found a better candidate than in previous work for Pt$_3$O$_6$ after $\sim$650 steps, and additional new global minimum candidates for the Pt$_6$O$_5$ and Pt$_6$O$_4$ compositions after $\sim$1300 and $\sim$1550 steps, respectively.  

On the other hand, the worst execution targeting the $\mu_2$ shows an undesirable focus on stoichiometries different from the most stable one at the target environment, Pt$_3$O$_4$. The algorithm focuses instead on the Pt$_3$O$_2$ composition (red) and other less relevant stoichiometries (gray) interposed with some explorations of Pt$_3$O$_4$. Thus, the algorithm gets stuck at a minimum of Pt$_3$O$_2$ composition and never finds a structure with a Gibbs energy of formation equal to or lower than the one obtained with GOFEE in that environment, even after 3000 steps. The run has thus failed both to identify more stable minima for any of the compositions and to reach the target Gibbs energy of formation for the selected $\Delta\mu_O$ (dotted line). Most of the DFT evaluations in the run have been performed for candidates with the Pt$_3$O$_2$ composition.

In summary, multiple executions of the algorithm are successful in reproducing the results obtained for GOFEE, or even improving them. However, as shown in section \ref{sec1}, the algorithm operates best when a single stoichiometry is considerably more stable than others under the targeted environment. Otherwise, and as was the case for Ir$_3$O$_x$ clusters, the searches seem to be more strongly biased towards smaller, easier to optimize, stoichiometries.

\subsection{Performance analysis for the Pd(100) surface}

\begin{figure}[h]
    \centering
    \includegraphics[width=0.40\textwidth]{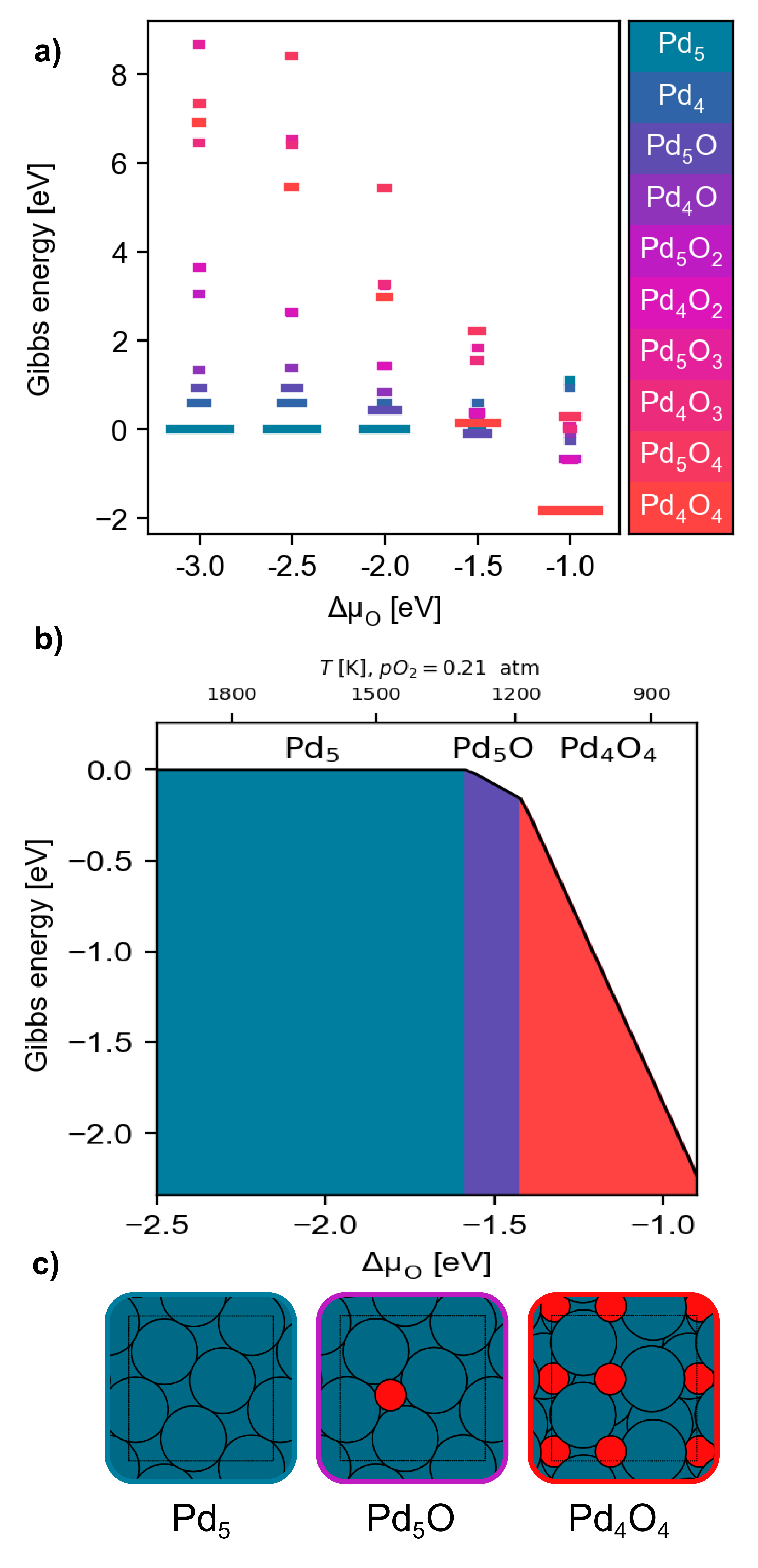}
    \caption{\textbf{a)} Gibbs energy of formation (with respect to the pristine surface, i.e., the Pd$_5$ state) of the best structure found with the GCGO algorithm for each stoichiometry among all oxidized states of the Pd(100)surface considered. GCGO executions have been carried out at five different $\Delta\mu_O$ values (-1.0, -1.5, -2.0, -2.5, and -3.0 eV). The width of 
    each bar is proportional to the number of evaluations the algorithm 
    has performed at the corresponding stoichiometry. \textbf{b)} $\Delta G$ vs $\Delta\mu_O$ diagram derived from the GCGO searches. In the upper axis, $\Delta\mu_O$ is evaluated in terms of temperature at constant pressure. \textbf{c)} Structures of the Pd$_5$, Pd$_5$O, and Pd$_4$O$_4$ phases.}
    \label{fig:pd100_search}
\end{figure}

As a last case example, we apply the GCGO algorithm to obtain stable oxidized phases of the Pd(100) surface. Fig. \ref{fig:pd100_search}a shows the Gibbs energies of the best candidates obtained, as a function of the targeted $\Delta\mu_O$ and of the stoichiometry of the structure. As expected, for low 
chemical potentials corresponding to oxygen-lean conditions, the pristine metallic surface is preferred. 
The width of the bars is Fig. \ref{fig:pd100_search}a is proportional to the number of 
evaluations for that stoichiometry, revealing that the algorithm indeed 
focuses more on the most reduced Pd$_5$ and Pd$_4$ for runs in which oxygen is relatively unavailable or scarce, where the Pd$_5$ state corresponds to the regular metal surface with 5 Pd atoms in each layer of $\sqrt{5} \times \sqrt{5}R27^\circ$ supercell. For higher $\Delta\mu_O$, the algorithm finds that the surface 
with adsorbed oxygen is preferred, with the algorithm's focus being more evenly distributed
across different stoichiometries. For the largest $\Delta\mu_O$ (-1.0 eV), the algorithm 
focuses most of the DFT-energy evaluations on oxidized states, correctly identifying the surface-oxide reconstruction (i.e., a thin Pd$_4$O$_4$ film) as the global minimum structure and stoichiometry.

These results demonstrate the 
algorithm's ability to focus on relevant oxidation state also of extended systems and to identify relevant phases at each targeted
chemical potential. Fig. \ref{fig:pd100_search}b  shows the phase diagram that can be constructed from the structures found by the GCGO algorithm, which is in excellent agreement with previous results \cite{Pd100-Rogal2007}.

\section{Conclusions}\label{concl}
In this work, we have presented a machine-learning assisted grand canonical global optimization algorithm, implemented as a module for the Atomistic Global Optimization X (AGOX) framework. We have evaluated the computational performance of the algorithm and demonstrated its ability to reproduce some examples from the literature.

The analysis of the performance has shown that the algorithm is able to find stable configurations and compositions for the targeted environmental conditions with a higher efficiency than Canonical approaches in which all of the considered stoichiometries are explored independently. GCGO algorithm runs are also routinely able to find the global minima of various stoichiometries during a single run, which allows revealing regions of the phase diagrams not only near the targeted conditions of interest, but also at other more or less reducing regions. 

The presented algorithm therefore constitutes a promising alternative to existing methods. We thus expect the GCGO algorithm to become a relevant tool in the characterization of solid catalysts under reaction conditions, where the composition and structure of the stable phases is not known \textit{a priori} and challenging to determine even with state-of-the-are experimental methods. This implementation also paves the way for further improvements in computational efficiency, which should allow to tackle larger, and therefore more complex, nanostructured materials and their reconstructions in reactive environments. 

\section{Supplementary Material}
The supplementary material file contains additional figures S1, S2, S3, and S4. \hl{Figure S1 includes p,T phase diagrams showing stable stoichiometries for Ir$_3$O$_x$ clusters using different approximations. Figures S2, S3, and S4 illustrate the training costs and accuracy of different MLIPs.}

\section{Acknowledgements}\label{cack}
This work was supported by the Spanish/FEDER Ministerio
de Ciencia, Innovación y Universidades [Grant Nos. PID2021-
128217NB-I00, MDM-2017-0767, CEX2021-001202-M, PID2022-140120OA-I00, PID2024-157317NB-I00, and RYC2021-032281-I (for A.B.)] as well as by the Generalitat de Catalunya [Grant No.
and 2021SGR00286]. Computer resources have been partly
provided by the Red Española de Supercomputación. This study was also supported by the European COST Actions CA18234 and CA21101.
This work was also supported by the VILLUM FONDEN through Investigator grant, project no. 16562, and by the Danish National Research Foundation (Danmarks Grundforskningsfond) through the Center of Excellence “InterCat” (Grant agreement no: DNRF150).

\section{DATA AVAILABILITY}\label{data}
The implemented algorithms (along with documentation) are
available within the AGOX 3.11.0 python library (https://gitlab.com/agox/agox). The data that support the findings of this study are openly available in \hl{https://doi.org/10.5281/zenodo.16809151}, including: the most stable structures found during the Ir$_3$O$_x$, Pt$_3$O$_x$/CeO$_2$(111), and Pd(100) GCGO runs; the corresponding GCGO runs and GCGO execution scripts; the Ir$_3$O$_x$ MACE model and the NequIP and MACE models trained on the Pt$_3$O$_x$/CeO$_2$(111) structures for the comparison to the GPR+SOAP model\hl{; the GPR+SOAP models produced after each of the Pt$_3$O$_x$/CeO$_2$(111) GCGO runs and the python scripts used to generate them.}

%
%

%


\bibliography{refs/references}
\end{document}